\documentclass[prb,onecolumn,showpacs,preprintnumbers,amsmath,amssymb,superscriptadress,bibnotes,superscriptaddress]{revtex4}

\usepackage{color}
\usepackage{ulem}
\usepackage[dvipdfmx]{graphicx}
\usepackage{dcolumn}
\usepackage{bm}

\usepackage{array}
\usepackage{float}
\usepackage{units}
\usepackage{multirow}
\usepackage{amsmath}
\usepackage{setspace}

\usepackage{mathrsfs}

\usepackage{natbib}

\begin{document}


\title{Quantum Criticality in YbCu$_4$Ni}

\author{Kotaro Osato}
\affiliation{Institute for Materials Research, Tohoku University, Katahira, Sendai 980-8577, Japan}
\affiliation{Department of Physics, Tohoku University, Aoba, Sendai 980-8578, Japan}
\author{Takanori Taniguchi}
\email{Author to whom correspondence should be addressed: takanori.taniguchi.d3@tohoku.ac.jp}
\affiliation{Institute for Materials Research, Tohoku University, Katahira, Sendai 980-8577, Japan}
\email{takanori.taniguchi.d3@tohoku.ac.jp}
\author{Hirotaka Okabe}
\affiliation{Institute for Materials Research, Tohoku University, Katahira, Sendai 980-8577, Japan}
\author{Takafumi Kitazawa}
\author{Masahiro Kawamata}
\author{Zhao Hongfei}
\affiliation{Institute for Materials Research, Tohoku University, Katahira, Sendai 980-8577, Japan}
\affiliation{Department of Physics, Tohoku University, Aoba, Sendai 980-8578, Japan}
\author{Yoichi Ikeda}
\affiliation{Institute for Materials Research, Tohoku University, Katahira, Sendai 980-8577, Japan}
\author{Yusuke Nambu}
\affiliation{Institute for Materials Research, Tohoku University, Katahira, Sendai 980-8577, Japan}
\affiliation{Organization for Advanced Studies, Tohoku University, Sendai 980-8577, Japan}
\affiliation{FOREST, Japan Science and Technology Agency, Kawaguchi, Saitama 332-0012, Japan}
\author{Dita Puspita Sari}
\affiliation{Innovative Global Program, College of Engineering, Shibaura Institute of Technology, Saitama 337-8570, Japan}
\affiliation{Meson Science Laboratory, Nishina Center for Accelerator-Based Science, RIKEN, Wako, Saitama 351-0198, Japan}
\author{Isao Watanabe}
\affiliation{Meson Science Laboratory, Nishina Center for Accelerator-Based Science, RIKEN, Wako, Saitama 351-0198, Japan}
\author{Jumpei G Nakamura}
\affiliation{Institute of Materials Structure Science, High Energy Accelerator Research Organization, Tsukuba, Ibaraki 305-0801, Japan}
\author{Akihiro Koda}
\affiliation{Institute of Materials Structure Science, High Energy Accelerator Research Organization, Tsukuba, Ibaraki 305-0801, Japan}
\author{Jun Gouchi}
\affiliation{The Institute for Solid State Physics, University of Tokyo, Kashiwa, Chiba 277-8581, Japan}
\author{Yoshiya Uwatoko}
\affiliation{The Institute for Solid State Physics, University of Tokyo, Kashiwa, Chiba 277-8581, Japan}
\author{Masaki Fujita}
\affiliation{Institute for Materials Research, Tohoku University, Katahira, Sendai 980-8577, Japan}

\date{\today}

\begin{abstract}

We report on the quantum criticality of YbCu$_4$Ni as revealed by our combined micro- and macro-measurements. We determine the crystal structure of YbCu$_4$Ni with site mixing by neutron diffraction measurements, which suggests the possible presence of Kondo disorder. However, decreasing the local spin susceptibility distribution and the development of spin fluctuations below 10 K at ambient pressure by muon spin rotation and relaxation measurements suggests that YbCu$_4$Ni exhibits quantum criticality. Therefore, our experimental results indicate that YbCu$_4$Ni is a new material that exhibits quantum criticality under a zero magnetic field and ambient pressure.

\end{abstract}


\maketitle

\section{\label{sec:level1}Introduction}
Quantum criticality, which induces exotic physical properties due to spin fluctuations, is one of the most important research topics in solid-state physics. 
The unique properties, such as an unconventional superconductivity, spin liquids, and other unique physical properties emerge due to spin fluctuations~\cite{Stewart_1984, Brando_2016, Zhou_2017, Subir_2023}, near the quantum critical point (QCP).
At QCP, a secondary phase transition occurs at absolute zero temperature and spin fluctuation develop due to the suppression of the entropy releasing.
Currently, research on these phenomena and their origin has gained attraction, but high magnetic fields and pressure is limiting the development of the research.
In this paper, our results find that YbCu$_4$Ni exhibits quantum criticality under a zero magnetic field and ambient pressure.

In $f$ electron system, $c$-$f$ hybridization provides two interactions, Ruderman-Kittel-Kasuya-Yosida (RKKY) interaction and Kondo effect~\cite{Doniach_1977, Coleman_2001, Gegenwart_2008, Si_2010}. 
RKKY interaction stabilizes the magnetic order. 
On the other hand, Kondo effect screens the magnetic moment of the $f$ electrons.
The QCP appears at the case that the energy scales of RKKY interaction and Kondo effect are the same.
Many anomalous physical properties appear in the vicinity of the QCP because of spin fluctuations. 
In particular, large electronic specific heat coefficients are detected near the QCP, e.g., in CeCu$_6$~\cite{OTT_1985}, CeAl$_3$~\cite{Andres_1975}, and YbRh$_2$Si$_2$~\cite{Gegenwart_2002}.
By contrast, large electronic specific heat coefficients can also originate from the Kondo disorder effect, in which various disorders (such as those in a crystal structure) result in a widely distributed Kondo temperature $T_{\mathrm{k}}$. 
However, to date, Kondo disorder and quantum criticality have been experimentally distinguished in only a few systems. 
For example, UCu$_4$Pd exhibits several unusual physical properties such as a large electronic specific heat coefficient and an electrical resistivity with a non-Fermi liquid behavior~\cite{Otop_2005, Weber_2001, Giudicelli_2005} due to the Kondo disorder.
Furthermore, nuclear magnetic resonance (NMR)~\cite{Bernal_1995} and muon spin rotation and relaxation ($\mu$SR)~\cite{MacLaughlin_1998, MacLaughlin_2000} spectroscopic analyses reveal an inhomogeneous distribution of the local spin susceptibility.
These non-Fermi liquid behavior and inhomogeneous distribution can be explained by  the case of a widely distributed Kondo temperature.
Neutron diffraction~\cite{Chau_1998} and X-ray absorption fine structure (EXAFS)~\cite{Booth_1998} measurements reveal that the $T_{\mathrm{k}}$ distribution shown by this material originates from the disorder of Cu sites.
Therefore, these previous studies on UCu$_4$Pd revealed that determining the distribution of the local susceptibility and crystal structure is a powerful method for experimentally distinguishing between the quantum criticality and Kondo disorder behavior.

In a recent study, YbCu$_4$Ni exhibited a large electronic specific heat coefficient ($\gamma _e \sim 7.5$  J/mol K$^2$)~\cite{Sereni_2018}.
Because of the adjacent atomic number of Cu and Ni, YbCu$_4$Ni may show site mixing similar to UCu$_4$Pd, and thus, Kondo disorder due to crystal disorders is expected in this system. 
On the other hand, because YbCu$_5$, which is close to QCP, exhibits a large electronic specific heat coefficient of 0.55 J/mol K$^2$~\cite{Tsuji_1997}, YbCu$_4$Ni, whose lattice constant is similar to that of YbCu$_5$, is also expected to be closer to QCP. 
Therefore, there are two possible origins for the reported large $\gamma _e$ value~\cite{Sereni_2018} of YbCu$_4$Ni, either Kondo disorder or quantum criticality. 
To determine the origin of the large $\gamma _e$ value, we performed neutron powder diffraction (NPD) and $\mu$SR measurements, which are powerful microscopic techniques that reveals the distribution of local spin susceptibility and crystal structure, same as the previous studies on UCu$_4$Pd. 
A comprehensive analysis of the results revealed the existence of quantum criticality in YbCu$_4$Ni.

\section{\label{sec:level2}Experimental Details}
In this study, we prepared poly-crystalline YbCu$_4$Ni samples by arc melting.
In this process, ingots of Yb (99.9\% purity), shots of Cu (99.99\% purity), and Ni (99.99\% purity) were  mixed in a ratio of 1.1:4.0:1.0 (Yb:Cu:Ni).
The sample weighed approximately 4 g and was sealed in a quartz tube with argon gas and annealed at 800${}^\circ$C for two weeks.
The sample preparation process was the same as that applied in the previous study, in which wavelength-resolved electron beam analysis performed at Hiroshima University confirmed that the samples were composed of Yb, Cu, and Ni in the ratio of 1:4:1 (Yb:Cu:Ni)~\cite{Shimura_2022}.
In the present study, the phase purity of YbCu$_4$Ni was examined by X-ray powder diffraction (XRD) and NPD measurements.
In the XRD measurements, Cu-$K_\alpha$ radiation with a wavelength ($\lambda$) of 1.54 \AA $ $ was used to collect the diffraction profiles. 
In addition, we conducted neutron diffraction measurements using Ge(331) and Ge(311) reflection ($\lambda = 2.20$ and $2.01$ \AA) on HERMES and AKANE diffractometers at Japan Research Reactor No. 3 (JRR-3), Japan.
The electrical resistivity of the sample was measured by the four-probe method using an ac resistance bridge in a standard configuration; the measurements above and below 2 K were conducted using a Quantum Design physical property measurement system and a dilution refrigerator, respectively.
 The $\mu$SR measurements were performed using the S1 (ARTEMIS) and D1 $\mu$SR spectrometer, equipped with a $^3$He cryostat and a dilution refrigerator at Materials and Life Science Experimental Facility (MLF) in J-PARC, Japan, and the spectrometer (ARGUS) at Port 2 in the RIKEN-RAL Muon Facility in the Rutherford Appleton Laboratory (RAL), UK.

\begin{table*}[htb]
\renewcommand{\arraystretch}{1.5}  
\centering
 \caption{Structural parameters of YbCu$_4$Ni refined from neutron powder diffraction at $T=3$ K.}
\label{table:npd}
\begin{tabular}{ccccccc||ccccc}
\hline 
\multicolumn{7}{c||}{ordered model (a)} & \multicolumn{5}{c}{disordered model (b)}\tabularnewline
\hline 
\hline 
atom & site & x & y & z & $B_{\mathrm{iso}}$ & occ. (\%) & x & y & z & $B_{\mathrm{iso}}$ & occ. (\%)\tabularnewline
\hline 
\hline 
Yb & 4a & 0 & 0 & 0 & 0.432(5) & 100 & 0 & 0 & 0 & 0.421(1) & 100\tabularnewline
\hline 
Ni & \multirow{2}{*}{4c} & \multirow{2}{*}{$\frac{1}{4}$} & \multirow{2}{*}{$\frac{1}{4}$} & \multirow{2}{*}{$\frac{1}{4}$} & \multirow{2}{*}{1} & 100  & \multirow{2}{*}{$\frac{1}{4}$} & \multirow{2}{*}{$\frac{1}{4}$} & \multirow{2}{*}{$\frac{1}{4}$} & \multirow{2}{*}{0.406(7)} & 0\tabularnewline
Cu &  &  &  &  &  & 0 &  &  &  &  & 100\tabularnewline
\hline 
Ni & \multirow{2}{*}{16e} & \multirow{2}{*}{0.624(2)} & \multirow{2}{*}{0.624(2)} & \multirow{2}{*}{0.624(2)} & \multirow{2}{*}{0} & 0 & \multirow{2}{*}{0.624(3)} & \multirow{2}{*}{0.624(3)} & \multirow{2}{*}{0.624(3)} & \multirow{2}{*}{0.268(8)} & 25\tabularnewline
Cu &  &  &  &  &  & 100 &  &  &  &  & 75\tabularnewline
\hline 
\hline 
$R_{\mathrm{p}}$: 5.38 &$R_{\mathrm{wp}}$: 8.24  &$\chi^{2}$: 4.02  &  &  &  &  & $R_{\mathrm{p}}$: 5.10 & $R_{\mathrm{wp}}$: 7.23 & $\chi^{2}$: 3.10 &  & \tabularnewline
\hline 
\end{tabular}
\end{table*}

\begin{figure}[h]
\vspace*{10pt}
\begin{center}
\includegraphics[width=8.5cm,clip]{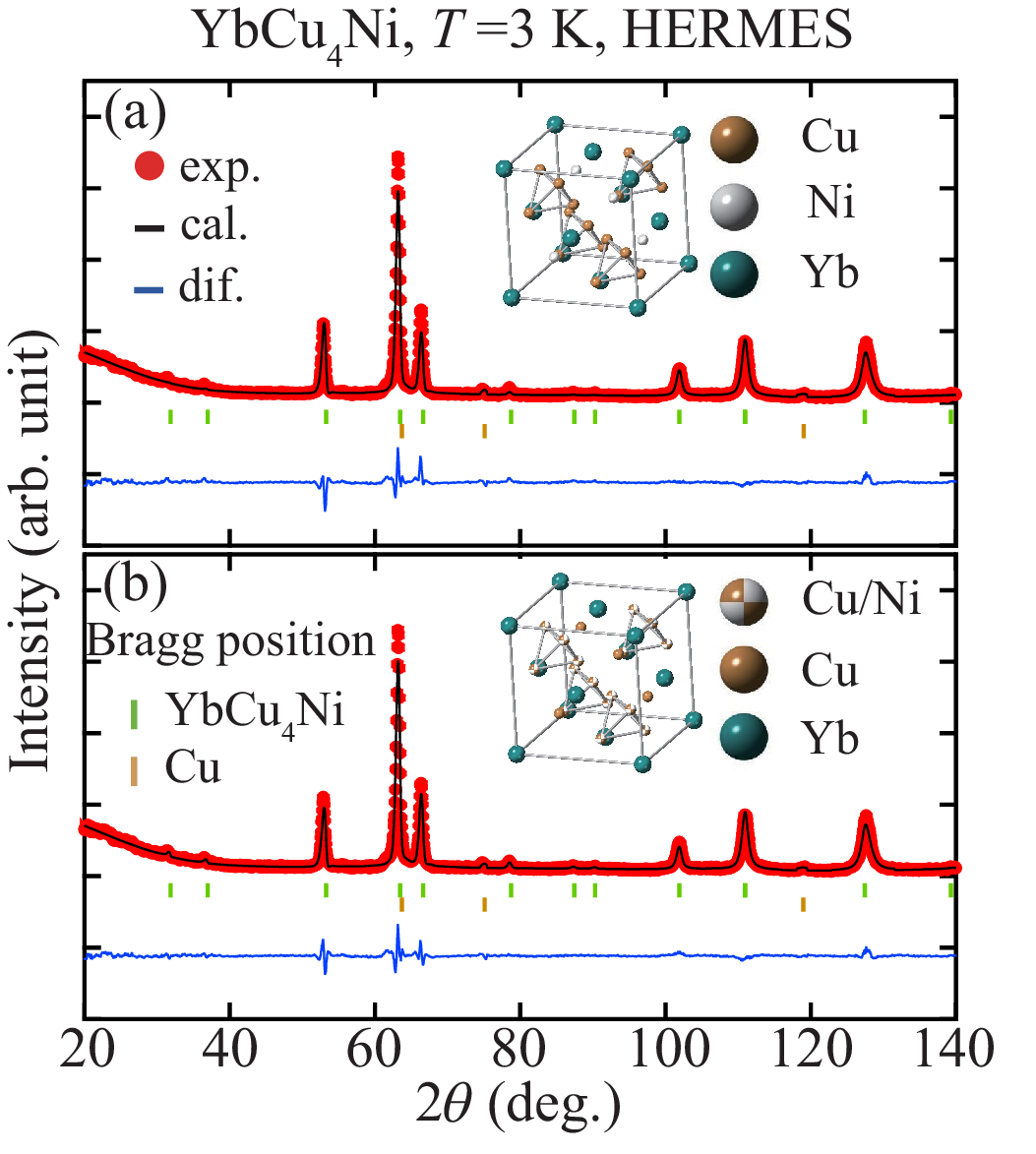}
\end{center}
\caption{(color online) (a) NPD profiles of YbCu$_4$Ni refined using the model (a). (b) NPD profiles of YbCu$_4$Ni refined using model (b). Insets represent the crystal structures derived using ordered model (a) and disordered model (b).
}
\label{Fig.1}
\end{figure}
\section{\label{sec:level3}Results}
\subsection{\label{sec:level3_1}Neutron Powder Diffraction}

The NPD profiles of the samples were recorded at 3 K with a minimal thermal contribution to determine their crystal structures at HERMES. 
Because the atomic numbers of Cu and Ni are adjacent, two crystal structures are possible, which are defined as ordered model (a) and disordered model (b) as shown in the insets in Figs. 1(a) and 1(b); in both the cases, $F$$\bar{4}$3$m$ (No. 216).
Model (a) shows a the AuBe$_5$-type structure, with Wyckoff positions 4a, 16e, and 4c for Yb, Cu, and Ni, respectively.
Model (b) exibits Cu and Ni site mixing at 4c and 16e. 
The crystal structures were determined by Rietveld analysis, which was performed using the FULLPROF package~\cite{Rodriguez_1993}, and the corresponding results are shown in Figs. 1(a) and 1(b) and listed in Table I. 
Both models are capable of explaining the position of the signal. 
Moreover, the result of NPD at HERMES is consistent with the results of XRD and NPD at AKANE. 
The details about experimental results are in the Supplementary Materials.
$B_{\mathrm{iso}}$ is isotropic displacement parameter and the range was set to $0\leq B_{\mathrm{iso}}\leq1$.
In disordered model (b), where the occupancy was also treated as a free parameter, the occupancy of Cu in 4c becames 100\%.
Further, a comparing between the Rietveld discrepancy values, i.e., goodness-of-fit, $R_{\mathrm{p}}$, $R_{\mathrm{wp}}$, $\chi ^2$, of both the models reveals that the ability of disordered model (b) to reproduce the experimental results surpasses that of ordered model (a). 
Based on this result, we can conclude that disordered model (b) best describes the crystal structure of YbCu$_4$Ni.

\begin{figure}[h]
\vspace*{10pt}
\begin{center}
\includegraphics[width=8.5cm,clip]{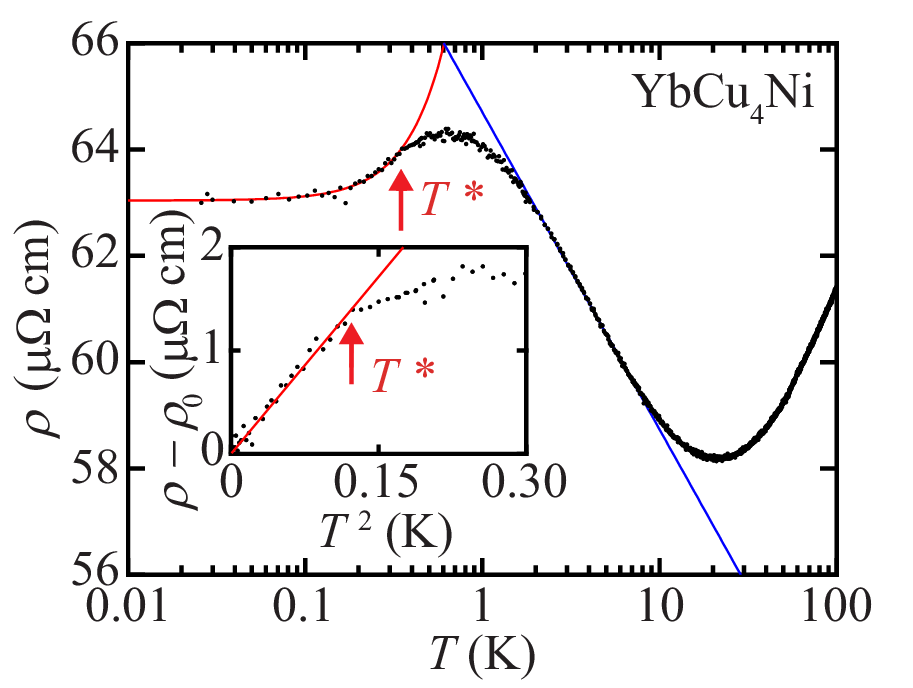}
\end{center}
\caption{(color online) Temperature dependence of the electrical resistivity of YbCu$_4$Ni. Inset represents $\rho$ versus $T^2$ at low temperatures. The red and blue lines represent the curves fitted with $\rho\left(T\right)=\rho_{0}+AT^{2}$ and $\rho\left(T\right)=C\ln\left(T/T_{0}\right)$.
}
\label{Fig.2}
\end{figure}
\subsection{\label{sec:level3_2}Electrical Resistivity}
Figure 2 shows the temperature dependence of the electrical resistivity.
In the range of 2 - 20 K, the electrical resistivity values were fitted by a function $-\ln\left(T/T_{0}\right)$. 
Below $T^* = 0.35$ K, the measured resistivity values were reproduced using the function $\rho_{0}+AT^{2}$, as shown in the inset of Fig. 2. 
The experimental datapoints were fitted by the function $\rho\left(T\right)=\rho_{0}+AT^{2}$, and $A = 11.43$ $\mu$$\Omega$ cm/K$^2$ was obtained, which indicated the formation of a heavy Fermi liquid state below $T^*$.

\begin{figure}[h]
\vspace*{10pt}
\begin{center}
\includegraphics[width=8.5cm,clip]{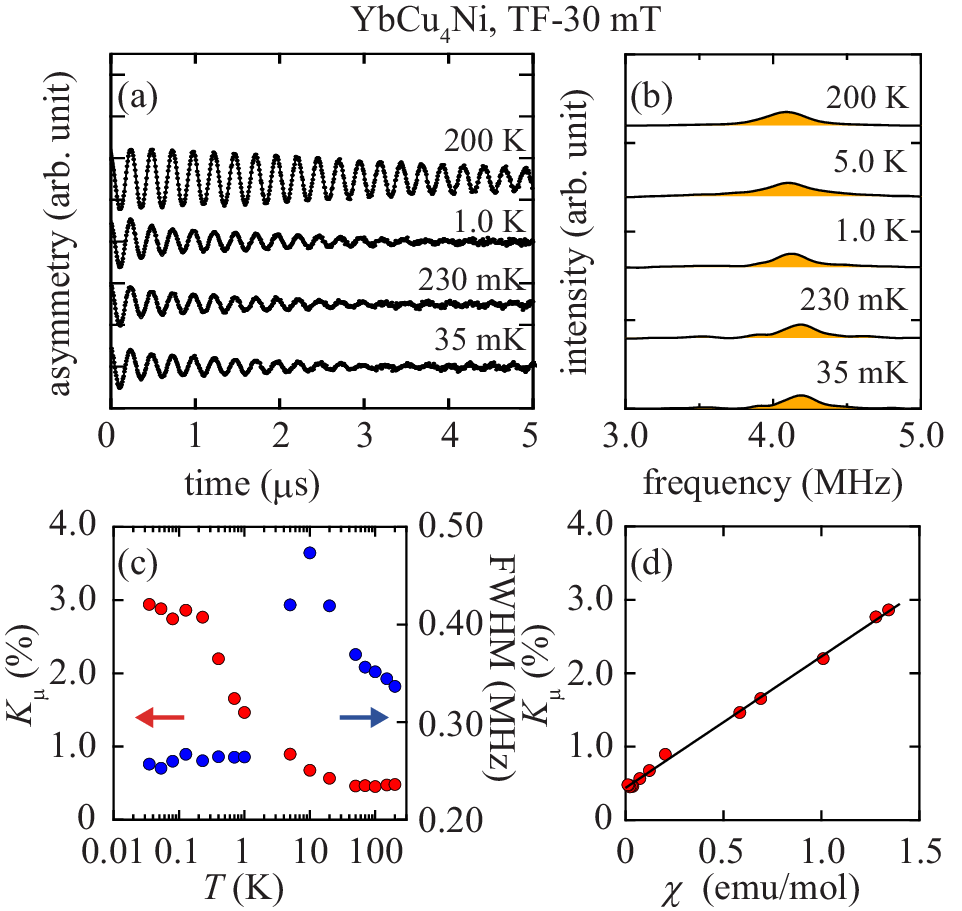}
\end{center}
\caption{(color online) (a) Temperature dependence of $\mu$SR time spectra of YbCu$_4$Ni under a TF of 30 mT. The data are analyzed using the MUSRFIT software pagckage~\cite{Suter_2012}. (b) Temperature dependence of FFT $\mu$SR spectra. (c) the temperature dependence of the muon Knight shift (red circle and FWHM (blue circle). (d) $K$-$\chi$ plot. $\chi$ is adopted from a previously reported study~\cite{Sereni_2018}. 
}
\label{Fig.3}
\end{figure}
\subsection{\label{sec:level3_3}TF-$\mu$SR}
Transverse magnetic field (TF)-$\mu$SR measurements were performed with D1 $\mu$SR spectrometer to microscopically investigate the behavior of the static spin susceptibility. 
Figure 3(a) shows the temperature dependence of $\mu$SR time spectra for TF of 30 mT. 
The observed increase in the relaxation can be ascribed to spin fluctuations, and no distinct changes are observed below 1 K. 
These observations are consistent with the longitudinal magnetic field (LF)-$\mu$SR measurement results described later in this paper. 
The spectrum obtained by fast Fourier transformation (FFT) of the time spectra is shown in Fig. 3(b), which also depicts the cell-derived background used in the experiment. 
No distinct splitting of the FFT spectrum is observed, indicating that no magnetic transition occurs in this system, which is consistent with the electrical resistivity measurement results. 
The temperature dependences of muon Knight shift ($K_\mu$) and full width at half maximum (FWHM), estimated by fitting the FFT spectrum using Lorentzian function, are displayed in Fig. 3(c). 
The muon Knight shift increases with decreasing temperature above the characteristic temperature $T^* = 0.35$ K and is constant below $T^*$, indicating that electron state is nonmagnetic. 
In addition, the line width is constant below $T^*$. 
These results collectively confirm the formation of a Fermi liquid state below $T^*$, mentioned previously. 
Figure 3(d) shows the $K_\mu$-$\chi$ plot above 130 mK. 
The Knight shift and magnetic susceptibility exhibit a linear relationship, indicating that muons do not destroy the $f$ electron state over the entire temperature range. 
The hyperfine coupling constant $A_{\mathrm{hf}}$ estimated from the slope of the $K_\mu$-$\chi$ plot is $A_{\mathrm{hf}} = 0.01$ T/$\mu_\mathrm{B}$.

\begin{figure}[h]
\vspace*{10pt}
\begin{center}
\includegraphics[width=8.5cm,clip]{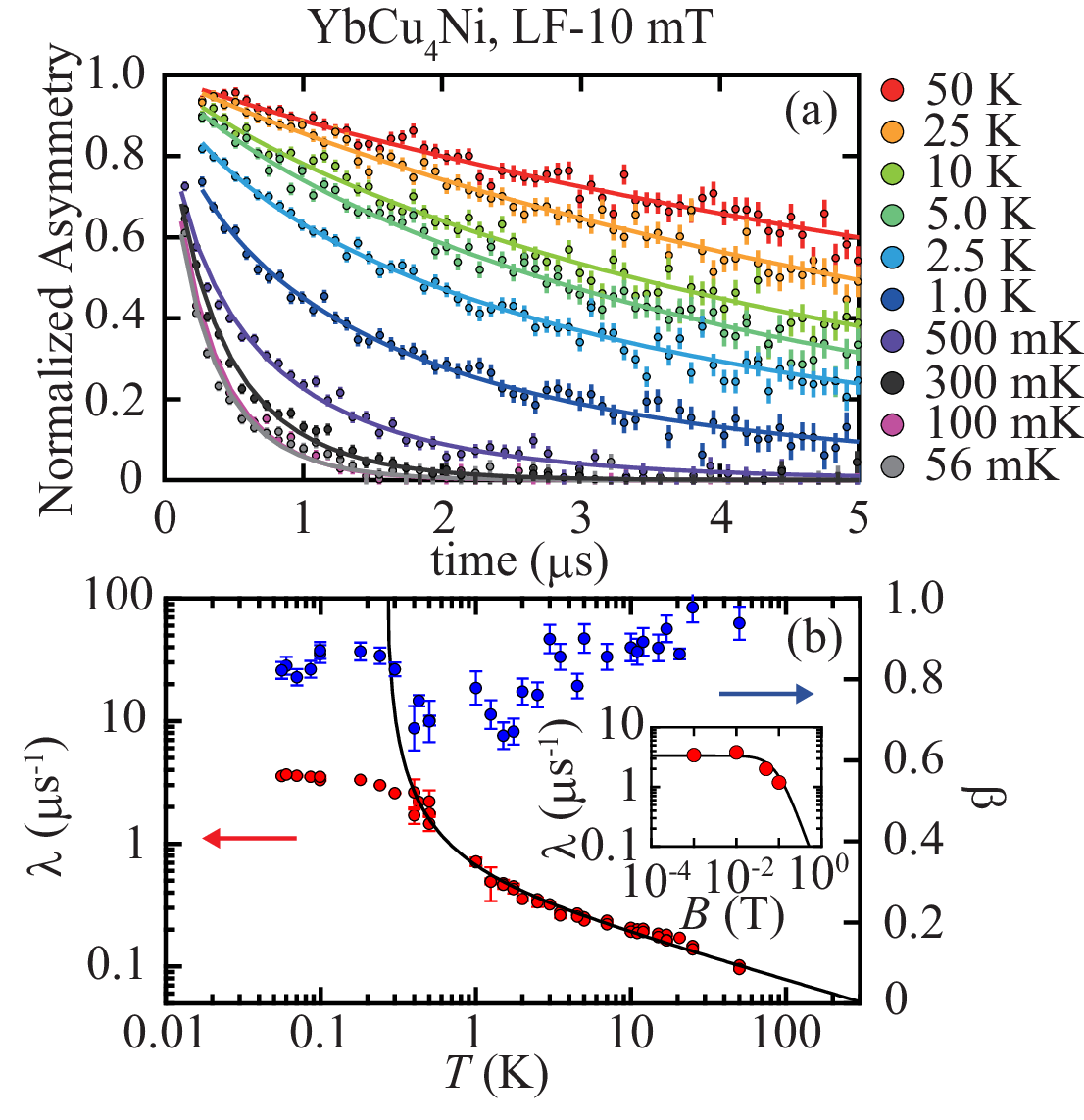}
\end{center}
\caption{(color online) (a) Temperature dependence of the $\mu$SR time spectra under LF of 10 mT. (b) Temperature dependence of the muon spin relaxion rate (red circle) and $\beta$ (blue circle). Solid line is fitting function: $\lambda\propto\frac{T}{\left(T-T_\mathrm{c}\right)^{1.38}}$. Inset represents the field dependence of the muon relaxation rate at 56 mK. Solid line is fitting function:  $\lambda\propto\left[1+\left(\frac{\mu_\mathrm{B}B}{k_\mathrm{B}T}\right)^{2}\right]^{-1}$.
}
\label{Fig.4}
\end{figure}
\subsection{\label{sec:level3_4}LF-$\mu$SR}
LF-$\mu$SR measurements were performed, with ARTEMIS, D1 $\mu$SR spectrometer, and ARGUS, to microscopically investigate the behavior of the dynamical spin susceptibility without the nuclear contributions. 
Figure 4(a) shows the temperature dependence of $\mu$SR time spectra under LF of 10 mT. 
Below 50 K, curves can be fitted by an exponential function, suggesting the presence of a fast depolarization component by electrons. 
From following four results we conclude that YbCu$_4$Ni does not exhibit a magnetic transition.
(i) No clear oscillations are observed at the lowest temperature (56 mK). (ii)  The intensity of the muon asymmetry is always constant. (iii) The relaxation rate does not decrease at low temperatures. (iv)  There is no evidence of a 1/3 tail of time spectra, typical static magnetism, found in the Zero-field and LF time spectra up to more than 10 microseconds.
These results are consistent with the measured electrical resistivity and TF-$\mu$SR results. 
To estimate the spin relaxation rate, the corresponding curves were fitted with a phenomenological equation, the stretched exponential function:
\begin{equation}
A(t) = A_0 e^{-(\lambda t)^\beta},
\end{equation}
where $\lambda$ is the muon spin relaxation rate. $A_0$ does not change with temperature, indicating that no phase separation occurs in the sample. 
The temperature dependences of $\lambda$ and $\beta$ is shown in Fig. 4(b). 
Above 0.3 K, $\lambda$ increases with decreasing temperature, implying the development of spin fluctuation. 
Below 0.3 K, $\lambda$ is constant, suggesting that large spin fluctuations exist above 56 mK. 
$\beta$ represents the inhomogeneity of the magnetic field at the muon stopping site~\cite{Adroja_2008}. 
The values were close to 1 in a region where there is little temperature dependency in the relaxation rate, suggesting that the inhomogeneous magnetic fields are small at the muon stopping site. This supports the results of the electrical resistivity measurements.

\section{\label{sec:level4}Discussions}
There are two possible reasons for the large electronic specific heat coefficients of YbCu$_4$Ni: (i) Kondo disorder and (ii) quantum criticality. 
From the results of NPD measurements, the possibility of Kondo disorder cannot be ruled out only by specific heat measurements because site mixing exists in YbCu$_4$Ni. 
This is because its crystal structure is similar to that of UCu$_{4}$Pd~\cite{Chau_1998, Booth_1998}, which exhibit Kondo disorder, and can be similarly discussed. 
However, the Kondo disorder scenario could not explain the temperature dependence of the FWHM of the FFT spectra. 
When the effect of Kondo disorder is dominant, the FWHM increases with increasing magnetic susceptibility~\cite{Bernal_1995, MacLaughlin_1998, MacLaughlin_2000, Ishida_2003}. 
Based on these results, we conclude that YbCu$_4$Ni is an interesting material that exhibits quantum criticality.

In general, the muon spin relaxation rate is proportional to the temperature in Fermi liquid state samples, according to the so-called Korringa-law~\cite{Abragam_1961}:
\begin{equation}
\lambda=\frac{\pi}{4}\hbar\gamma_{\mu}^{2}A_{\mathrm{hf}}^{2}\rho^{2}_\mathrm{e}\left(\varepsilon_\mathrm{F}\right)k_\mathrm{B}T,
\end{equation}
where $\gamma_\mu$ and $A_{\mathrm{hf}}=\frac{K_{\mu}}{\chi}N_\mathrm{A}\mu_\mathrm{B}$ are the muon gyromagnetic ratio and the hyperfine coupling constant, respectively. 
From the slope of $K_\mu$-$\chi$ plot (Fig. 3(d)), $A_{\mathrm{hf}}=0.01$ T/$\mu_B$ is estimated. $\rho_\mathrm{e}\left(\varepsilon_\mathrm{F}\right)$ is the density of states. For a 3-dimensional Fermi liquid, since the electronic specific heat coefficient is $\gamma_\mathrm{e}=\frac{\pi^{2}}{3}\rho_\mathrm{e}\left(\varepsilon_\mathrm{F}\right)k_\mathrm{B}$, the muon spin relaxation rate can be expressed as:
\begin{equation}
\lambda=\frac{9\hbar}{4\pi k_\mathrm{B}}A_{\mathrm{hf}}^{2}\gamma_{\mu}^{2}\gamma_\mathrm{e}^{2}T.
\end{equation}
When $\gamma_\mathrm{e} =7.5$ J/mol K$^2$~\cite{Sereni_2018}  is used, $\lambda=3.0\times10^{-2}\times T$ ($\mu$s$^{-1}$) can be estimated.  
Because of the extermely slow fluctuations, the Korringa-law cannot be observed in our experiments, and the spin fluctuation contribution was dominant.

\begin{figure}[h]
\vspace*{10pt}
\begin{center}
\includegraphics[width=8.5cm,clip]{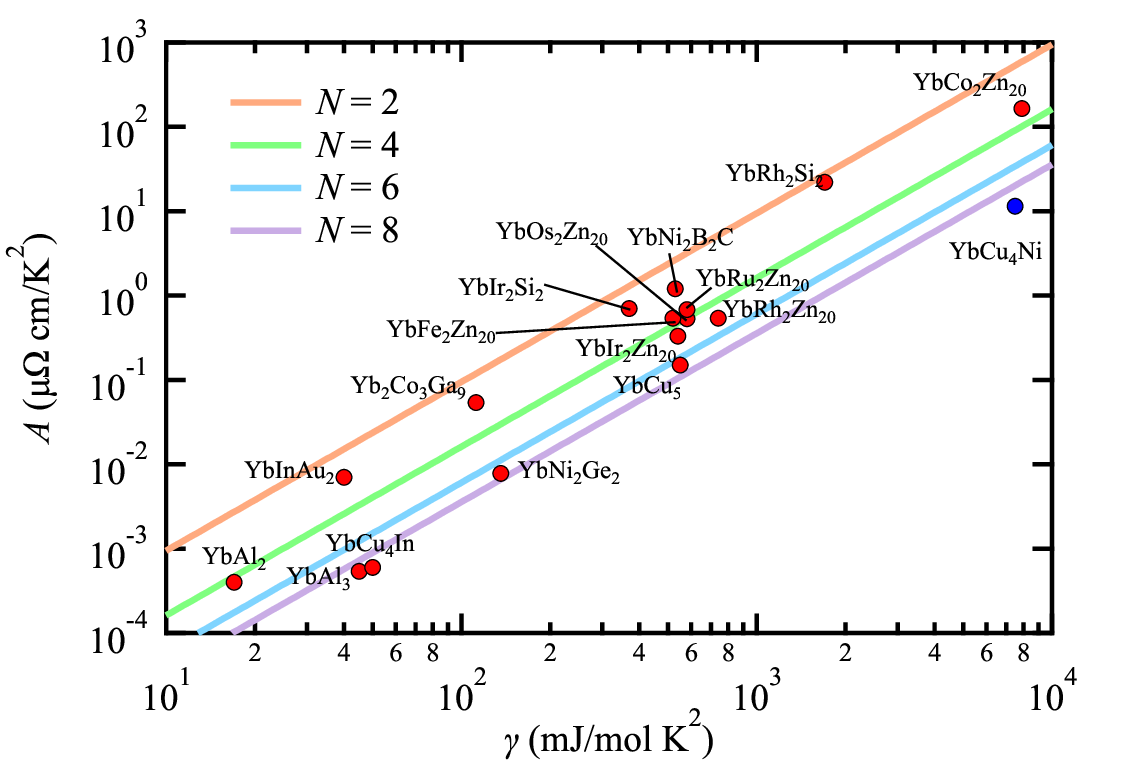}
\end{center}
\caption{(color online) Experimental values of the $T^2$-coefficient $A$ of resistivity and the $T$-linear coefficient of specific heat $\gamma$. Shown in this plot are the data presented by Kadowaki and Woods~\cite{Kadowaki_1986}, YbInAu$_2$~\cite{Tsujii_2003}, YbRh$_2$Si$_2$~\cite{Gegenwart_2002}, YbCo$_3$Ga$_9$~\cite{Dhar_1999}, YbAl$_2$~\cite{Tsujii_2003}, YbNi$_2$B$_2$C~\cite{Yatskar_1996}, Yb$T_2$Zn$_{20}$ ($T =$ Fe, Co, Ru, Rh, Os, Ir)~\cite{Torikachvili_2007}, YbIr$_2$Si$_2$~\cite{Hossain_2005}, YbAl$_3$~\cite{Ebihara_2000}, YbCu$_4$In~\cite{Tsujii_2003}, YbCu$_5$~\cite{Tsujii_2003}, YbNi$_2$Ge$_2$~\cite{Tsujii_2003}.
}
\label{Fig.5}
\end{figure}

 Above 0.3 K, the temperature dependence of the muon spin relaxation rate can be fitted using $\lambda\propto\frac{T}{\left(T-T_\mathrm{c}\right)^{1.38}}$  (Figure 4(b)).
This function corresponds to the case in which the two-dimensional ferromagnetic fluctuations in the self-consistent renormalization theory contribute to relaxation~\cite{Moriya_1985}.
Furthermore, the magnetic field dependence of the muon spin relaxation rate can be fitted using the function $\lambda\propto\left[1+\left(\frac{\mu_\mathrm{B}B}{k_\mathrm{B}T}\right)^{2}\right]^{-1}$, which is phenomenologically caused by the ferromagnetic fluctuations (inset of Fig. 4(b))~\cite{Carretta_2009}. 
On the other hand, the antiferromagnetic model could not reproduce the temperature and magnetic field dependence of $\lambda$.
We conclude that YbCu$_4$Ni is an interesting material with a ferromagnetic quantum criticality at a zero magnetic fields and ambient pressure.

Figure 5 shows the experimentally reported Kadowaki-Woods plot and our results, where $N$ is the ground state degeneracy. 
YbCu$_4$Ni is located on the $N=8$ line. 
If the disorder effect is significant, it deviates from this line, indicating that YbCu$_4$Ni exhibits quantum criticality~\cite{Tsujii_2003}. 
The unique properties of YbCu$_4$Ni arise from the Kondo effect and its quantum criticality at ambient pressure and a zero magnetic field. 
The correlation between the Kondo effect and quantum criticality with a large degeneracy will be explored in the future.

\section{\label{sec:level5}Conclusion}
In summary, YbCu$_4$Ni is a unique material that exhibits quantum criticality under a zero magnetic field and ambient pressure. 
NPD diffraction profiles determined the crystal structure with site-mixing, suggesting that Kondo disorder can occur. 
However, the decrease in the local spin susceptibility distribution below 10 K ruled out the Kondo disorder revealed by $\mu$SR measurements. 
The development of spin susceptibility below 10 K indicates that YbCu$_4$Ni is a quantum critical material.

\section*{Acknowledgements}
We thank K. Ishida, S. Kitagawa, T. Onimaru, T. Takabatake, Y. Kusanose, and Y. Shimura for stimulating discussions, and M. Ohkawara for his technical support at HERMES. The neutron diffraction experiment at JRR-3 was carried out under the general user program managed by the Institute for Solid State Physics, the University of Tokyo (Proposal Nos. 22410, 22616, and 23409), and supported by the Center of Neutron Science for Advanced Materials, Institute for Materials Research, Tohoku University. The $\mu$SR measurement at the Materials and Life Science Experimental Facility of J-PARC (Proposal Nos. 2020A0131, 2021B0402, and 2022A0064), and RIKEN-RAL Muon Facility in the Rutherford Appleton Laboratory (Experiment No. RB2070002) were performed under user program. This work was financially supported by JSPS/MEXT Grants-in-Aids for Scientific Research (Grant Nos. 19K23417, 21K13870, 23K13051).

\bibliographystyle{apsrev4-1}
\bibliography{YbCu4Ni}

\end{document}